\documentclass[twocolumn,showpacs,pra,aps]{revtex4}
\usepackage{ulem}
\usepackage{array}
\usepackage{amsmath}
\usepackage{graphicx}
\usepackage{dsfont}
\usepackage{color}
\def\bra#1{\langle #1|}
\def\ket#1{|#1 \rangle}

\begin{document}

\title{High-dimensional Bell test for a continuous variable state in phase space and its robustness to 
detection inefficiency}

\author{Seung-Woo Lee and Hyunseok Jeong}

\affiliation{Center for Macroscopic Quantum Control, Department of
Physics and Astronomy, Seoul National University, Seoul, 151-742,
Korea}

\date{\today}

\begin{abstract}
We propose a scheme for testing high-dimensional Bell inequalities
in phase space. High-dimensional Bell inequalities can be recast
into the forms of a phase-space version using quasiprobability
functions with the complex-valued order parameter. We investigate
their violations for two-mode squeezed states while increasing the
dimension of measurement outcomes, and finally show the robustness
of high-dimensional tests to detection inefficiency.

\end{abstract}


\maketitle

\section{Introduction}
\label{section:Intro}

Quantum nonlocality confirms the validity of quantum mechanics
against the local-realistic theories by violations of the
constraints on the correlation between local measurement outcomes.
Such constraints, called Bell inequalities, were first proposed by
Bell \cite{Bell64}, and to date, many versions have been proposed
and investigated \cite{CHSH69,Collins02,WSon05}. Probably the best
known version of Bell inequalities is Clauser, Horne, Shimony, and
Holt's (CHSH) inequality \cite{CHSH69}, which has been used for
verifying nonlocal correlations in a two-dimensional Hilbert space.
However, since most physical systems are composed of many particles
with many degrees of freedom and thus exhibit their properties in a
higher-dimensional Hilbert space, studying high-dimensional quantum
correlation is essential. Recently, orbital angular momentum states
of photon pairs \cite{Mair2001} and hyperentangled states
\cite{Barreiro2005} have been of great interest. It has been shown
that high-dimensional versions of quantum information processing
offer some advantages e.g. a robust quantum key distribution
\cite{Bruss2002}, superdense coding \cite{Barreiro2008}, and fast
high fidelity quantum computation \cite{Vallone2008}.

Several types of high-dimensional Bell inequalities have been
proposed and investigated in various ways. For example, the type
proposed by Collins {\em et al.} \cite{Collins02} is in the form of
a combination of joint probabilities, which we will call the CGLMP
inequality throughout this paper. The violation of the CGLMP
inequality was demonstrated for arbitrary high-dimensional entangled
states and experimentally realized for three-dimensional systems
\cite{Vaziri2002}. The type proposed by Son {\em et al.}
\cite{WSon05}, which we will call the SLK inequality, is in the form
of a combination of correlation functions. A generalized structure
of high-dimensional Bell inequalities was formulated both in joint
probability and correlation function representation where two
representations are in Fourier transform relations \cite{SWLee07}.

Phase-space formalism has been used successfully for describing
various quantum properties (especially for optical quantum states),
since any quantum state can be perfectly characterized by
quasiprobability functions such as the Wigner function, the
$Q$-function, and the $P$-function \cite{Cahill69}. Bell
inequalities in the forms of the CHSH inequality and the CH
inequality (another version of two-dimensional Bell inequality
\cite{CH74}) were proposed by the Wigner and $Q$-functions,
respectively \cite{Banaszek99}. Recently, a generalized version
merging the CHSH and CH types was formulated, which provides a way
of testing quantum nonlocality using quasiprobability functions with
an arbitrary nonpositive order parameter (which includes the
$Q$-function and the Wigner function) \cite{SWLee09}. However, so
far high-dimensional quantum nonlocality has been rarely studied in
phase-space formalism, in spite of a recent study \cite{WSon06},
probably because of the difficulty in discriminating multi-level
outcomes efficiently.

Indeed, the inefficiency of realistic detectors is one of the
biggest problems when implementing a Bell inequality test for
optical quantum states. The lowest efficiency bound for observing
the violation of local realism free from the detection loophole is
known to be very high (e.g. about 83\% for a Bell-CHSH inequality
test using an entangled photon pair), and such a high efficiency is
extremely difficult to achieve using current technology. It was
shown that the $Q$-function permits the lowest bound efficiency for
observing nonlocality in phase space \cite{SWLee09}. An entanglement
witness was proposed in phase space, which enables detecting
entanglement (but not nonlocality) even with significantly low
detection efficiencies \cite{SWLee10}. Very recently, it was shown
that high-dimensional Bell tests provide a lower bound for
detection-loophole-free nonlocality tests \cite{Tamas2010}.

In this paper we present a scheme for testing high-dimensional Bell
inequalities in phase-space formalism and show their robustness to
detection inefficiencies. The CHSH ineq,uality can be tested in
phase space \cite{Banaszek99} exploiting the fact that the Wigner
function is given as an expectation of the parity measurement on
photon number outcomes, i.e., $W(0)=(2/\pi)\langle (-1)^{\hat{n}}
\rangle$. This is a projection of photon number statistics of a
given quantum state to the two-dimensional Hilbert space with two
outcomes, $+1$ and $-1$. In our approach we increase the number of
outcomes to an arbitrary number $d$ by mapping the photon number
into a discrete phase in polar representation, and thus, the
outcomes are given as a complex variable $\langle \omega^{\hat{n}}
\rangle$, where $\omega = \exp({2\pi i/d})$. The expectation value
of $d$-level outcomes can be regarded as a generalized
quasiprobability function with a complex order parameter. This
approach has been already used in \cite{WSon06} to demonstrate the
violation of the CGLMP inequality for two-mode squeezed vacuum
states. In this paper, we (i) reformulate the CGLMP and SLK
inequalities in the forms of generalized structure using
quasiprobability functions, (ii) investigate their violations for
two-mode squeezed vacuum states with different numbers of outcomes,
and (iii) finally show that the CGLMP inequality can offer more
robust nonlocality tests to detection inefficiency than the CHSH
inequality.

This paper is organized as follows. In Sec.~\ref{section:HDBI} we
reformulate two types of high-dimensional Bell inequalities, CGLMP
and SLK, in the complex variable representation. We then investigate
their violations for two-mode squeezed vacuum states in
Sec.~\ref{section:VHBICV} and compare their tendencies as the number
of measurement outcomes $d$ increases. We also investigate the
effect of detection inefficiencies on the violations of
high-dimensional Bell inequalities by comparing it to the
two-dimensional case. Finally, we discuss and conclude our study in
Sec.~\ref{section:Remarks}.

\section{High-dimensional Bell inequalities in the complex-variable representation}
\label{section:HDBI}

In this section, we reformulate two types of high-dimensional Bell
inequalities, CGLMP and SLK, in the complex-variable representation.
Suppose that each observer independently chooses one of two
observables, $A_1$ or $A_2$ for Alice and $B_1$ or $B_2$ for Bob
with outcomes $k$ for Alice and outcomes $l$ for Bob, where $k,l \in
\{0,1,...,N-1\}$. Outcomes of each observable are binned to $d$
subsets by assigning complex variables $\omega^k$ and $\omega^l$,
where $\omega=\exp(2\pi i/d)$. We can define a correlation function
based on complex variables and then rewrite two types of Bell
inequalities in the complex-variable representation.

\subsection{Correlation functions mapped to complex variable}
\label{section:GCF}

A correlation function of two separately measured outcomes is
generally given in the form
\begin{eqnarray}
C=\sum^{N-1}_{k,l=0}\mu(k,l)P(A=k,B=l),
\end{eqnarray}
where $P(A=k,B=l)$ is the joint probability of Alice and Bob
obtaining outcomes $k$ and $l$ and $\mu(k.l)$ is the correlation
weight as a function of outcomes $k$ and $l$. We assume here that
the correlation weight $\mu(k,l)$ satisfies certain conditions
\cite{SWLee07}: (C1) The correlation expectation vanishes for a
bipartite system with a locally unpolarized subsystem,
$\sum_{k}\mu(k,l)=0, \forall l$ and $\sum_{l}\mu(k,l)=0, \forall k$.
(C2) The correlation weight is unbiased over possible outcomes of
each subsystem (i.e., translational symmetry within modulo $d$),
$\mu(k,l)=\mu(k+\gamma,l+\gamma), \forall \gamma$. (C3) The
correlation weight is uniformly distributed modulo $d$,
$|\mu(k+1,l)-\mu(k,l)|=|\mu(k,l+1)-\mu(k,l)|, \forall k,l$. These
are naturally required conditions for a symmetrical and locally
unbiased nature assigned to the correlation functions.

A correlation weight $\omega^{k-l}$ satisfies all these conditions
(though it is not a unique type), which is obtained by extending
correlation functions to complex variables. Higher-order ($n$)
correlations are represented by the $n$-th power of correlation
weight, $\omega^{n(k-l)}$ where $n$ is a positive integer. Thus, the
$n$-th order correlation function is
\begin{eqnarray} \label{eq:HOC}
C^{(n)}=\sum_{k,l=0}^{N-1}\omega^{n(k-l)}P(A=k,B=l),
\end{eqnarray}
which shows the periodicity of $C^{(d+n)}=C^{(n)}$. Note that any
Hermitian observable operator $\hat{H}$ can be associated with a
unitary operator $\hat{U}$ by the simple correspondence
$\hat{U}=\exp{(i\hat{H})}$. Therefore, any $N$-dimensional outcomes
of $A$ and $B$ can be mapped into complex values $\omega^{k}$ and
$\omega^{l}$ with a given $d$.

\subsection{CGLMP inequality}

We can reformulate the CGLMP inequality in complex variable
representation. The CGLMP function was originally proposed as a
combination of joint probabilities \cite{Collins02} and can be
written in a generalized form \cite{SWLee07}:
\begin{eqnarray}
\label{eq:CGLMP1} {\cal B}=
\sum^{2}_{a,b=1}\sum_{k',l'=0}^{d-1}\epsilon_{ab}(k',l')
P(\omega^{A_a}=\omega^{k'},\omega^{B_b}=\omega^{l'}),
\end{eqnarray}
with coefficients
\begin{eqnarray}
\label{eq:Coeff1} \nonumber
\epsilon_{11}(k',l')=1-\frac{2\dot{(k'-l')}}{d-1},~~
\epsilon_{12}(k',l')=1-\frac{2\dot{(l'-k')}}{d-1},\\
\nonumber \epsilon_{21}(k',l')=-1+\frac{2\dot{(l'-k')}}{d-1},~~
\epsilon_{22}(k',l')=1-\frac{2\dot{(k'-l')}}{d-1},
\end{eqnarray}
where the overdot implies the positive residue modulo $d$. The Bell
function (\ref{eq:CGLMP1}) should be bounded by $2$ in
local-realistic theories. The joint probability
$P(\omega^{A}=\omega^{k'},\omega^{B}=\omega^{l'})$ indicates the
probability that the outcomes by positive residue modulo $d$ of $A$
and $B$ are equal to $k'$ and $l'$, respectively. This is the
expectation of the projection operator
$\sum^{m-1}_{p=0}\ket{pd+k'}\bra{pd+k'}\otimes\sum^{m-1}_{q=0}\ket{qd+l'}\bra{qd+l'}$
in $d$-dimensional Hilbert space where we assume that $N=dm$ and $m$
is an integer.

We can rewrite the CGLMP function in terms of the correlation
functions (\ref{eq:HOC}). On the basis of the generalized formalism
in \cite{SWLee07}, any Bell type inequality can be written by the
sum of high-order correlation functions $C^{(n)}$ in complex space:
\begin{eqnarray}
\label{eq:GSBFc} {\cal B}=
\sum^{2}_{ab=1}\sum^{d-1}_{n=0}f_{ab}(n)C^{(n)}_{ab},
\end{eqnarray}
where the coefficients $f_{ab}(n)$ are functions of the correlation
order $n$ and the measurement configurations $a,b$. Note that it is
sufficient to consider first order to $d-1$-order correlation
functions due to the periodicity $C^{(d+n)}=C^{(n)}$. The
zeroth-order correlation has no meaning as it simply shifts the
value of ${\cal B}$ by a constant value and is thus here chosen to
vanish, i.e., $\sum_{a,b}f_{ab}(0)=0$. The CGLMP inequality can then
be recast into
\begin{eqnarray}
\label{eq:CGLMP2} \nonumber{\cal B}_{\mathrm{CGLMP}}&=&
\frac{2}{d-1}\sum_{n=1}^{d-1}\frac{1}{1-\omega^{-n}}\biggl[C^{(n)}_{11}-
\omega^{-n}C^{(n)}_{12}\\
&&~~~~~~~~~~~~~~~-C^{(n)}_{21} +C^{(n)}_{22}\biggr] \leq 2,
\end{eqnarray}
where
$C^{(n)}_{ab}=\sum^{d-1}_{k',l'=0}\omega^{n(k'-l')}P(\omega^{A_a}
=\omega^{k'},\omega^{B_b}=\omega^{l'})$ is the $n$-th order
correlation function. Note that the expectation value of the Bell
function in Eq.~(\ref{eq:CGLMP2}) is always real even though it is
represented by complex variables. The correlation function can be
obtained as an expectation value of the correlation operator,
\begin{eqnarray}
\label{eq:corrop} \nonumber
\hat{C}^{(n)}_{ab}&=&\sum^{d-1}_{k',l'=0}\omega^{n(k'-l')}\sum^{m-1}_{p=0}
\ket{pd+k'}_{a}\bra{pd+k'}\\
\nonumber
&&~~~~~~~~~~~~~\otimes\sum^{m-1}_{q=0}\ket{qd+l'}_{b}\bra{qd+l'},\\
&=&\sum^{N-1}_{k,l=0}\omega^{n(k-l)}\ket{k}_{a}\bra{k}\otimes\ket{l}_{b}\bra{l},
\end{eqnarray}
where each local measurement basis, denoted by the notation $a$, $b$
can be differentiated by unitary operation in $d$-dimensional
Hilbert space. Note that when $d=2$, Eq.~(\ref{eq:CGLMP2}) becomes
the CHSH inequality $C_{11}+C_{12}-C_{21}+C_{21}\leq2$ where
$\hat{C}_{ab}=\sum^{N-1}_{k,l=0}(-1)^{k+l}\ket{k}_a\bra{k}\otimes
\ket{l}_{b}\bra{l}$.

\subsection{SLK inequality}

We then consider the SLK inequality in the complex-variable
representation. On the basis of the generalized structure in
Eq.~(\ref{eq:GSBFc}), we can obtain the SLK function in the form of
Eq.~(\ref{eq:CGLMP1}) with the coefficients \cite{SWLee07}
\begin{eqnarray}
\nonumber \epsilon_{11}(k',l')=S(k'-l'+\frac{1}{4}),~~
\epsilon_{12}(k',l')=S(k'-l'-\frac{1}{4}),\\
\nonumber \epsilon_{21}(k',l')=S(k'-l'+\frac{3}{4}),~~
\epsilon_{22}(k',l')=S(k'-l'+\frac{1}{4}),
\end{eqnarray}
where $S(x\neq0)=(1/4)(\cot{\frac{\pi x}{d}}\sin{2\pi x}-\cos{2\pi
x}-1)$ and $S(x=0)=(d-1)/2$. The local-realistic bound of the SLK
function is given as a function of the number of outcomes $d$ by
$\frac{1}{4}(3\cot{\frac{\pi}{4d}}-\cot{\frac{3\pi}{4d}})-1$. In
order to compare it to the CGLMP inequality with a fixed
local-realistic bound $2$, we recast the original form of the SLK
inequality into
\begin{eqnarray}
\label{eq:SLK2} \nonumber {\cal B}_{\mathrm{SLK}} &=&
\frac{1}{R(d)}\sum_{n=1}^{d-1}\biggl[(\omega^{\frac{n}{4}}+\omega^{\frac{n-d}{4}})C^{(n)}_{11}\\
\nonumber
&&+(\omega^{-\frac{n}{4}}+\omega^{-\frac{n-d}{4}})C^{(n)}_{12}+(\omega^{\frac{3n}{4}}+
\omega^{\frac{3(n-d)}{4}})C^{(n)}_{21}\\
&&~~~~~~~~~~~~~~~~+(\omega^{\frac{n}{4}}+\omega^{\frac{n-d}{4}})C^{(n)}_{22}\biggr]
\leq 2,
\end{eqnarray}
where $R(d)=3\cot{\frac{\pi}{4d}}-\cot{\frac{3\pi}{4d}}-4$. Note
that the expectation value of Eq.~(\ref{eq:SLK2}) is always real and
when $d=2$ it becomes equivalent to the CHSH inequality.

Two high-dimensional Bell inequalities in the complex-variable
representation given in Eqs.~(\ref{eq:CGLMP2}) and (\ref{eq:SLK2})
can be effectively used for testing arbitrary $N$-dimensional
quantum states by arbitrary $d$-dimensional measurement. If we
consider the case $N=\infty$, we can perform high-dimensional Bell
tests for continuous-variable quantum states, as we will show in the
following section.

\section{Violations of high-dimensional Bell inequalities by
a continuous variable state} \label{section:VHBICV}

In this section, we investigate violations of two types of
high-dimensional Bell inequalities, CGLMP and SLK, for continuous
variable entangled states. We consider here the two-mode squeezed
vacuum states (TMSSs)
\begin{eqnarray}
 \label{eq:TMSVstate}
\ket{\Psi}_{\mathrm{TMSS}}=\sum_{j=0}^{\infty}\frac{\tanh^{j}{r}}{\cosh{r}}
\ket{j,j},
\end{eqnarray}
where $r>0$ is the squeezing parameter and $\ket{j}$ is the number
state of each mode. This can be realized by non degenerate optical
parametric amplifiers \cite{Reid88}, and highly entangled photon
pairs can be generated for testing Bell inequalities \cite{zhang07}.
Such states are well suited to Bell inequality tests since entangled
photon pairs can be generated and distributed over long distances
\cite{Weihs98,Tittel98}.

\subsection{Bell tests by reconstructing quasiprobability functions}
\label{section:BT}

Let us consider high-dimensional Bell tests by reconstructing
quasiprobability functions. An entangled state generated from a
source of correlated photons is distributed to two spatially
separated parties called Alice and Bob. Each party performs a local
measurement by counting photon numbers. The bases of each local
measurement are differentiated by the displacement operation
$\hat{D}(\alpha)$ for Alice and $\hat{D}(\beta)$ for Bob where
$\alpha$ and $\beta$ are complex variables associated with points in
phase space \cite{Cahill69}. If we bin the measured photon numbers
alternatively into two-dimensional outcomes ($+1$ and $-1$), the
expectation value of local measurement is given as a Wigner function
at the point displaced $\alpha$ (or $\beta$) in phase space, i.e.,
$W(\alpha)=(2/\pi)\langle (-1)^{\hat{n}(\alpha)} \rangle$, where
$\hat{n}(\alpha)=\hat{D}(\alpha)\hat{n}\hat{D}(-\alpha)$ is a
displaced number operator. A detail experimental setup for
reconstructing quasiprobability functions by photon counting is
given in \cite{Banaszek96}.

For high-dimensional outcomes, we bin the counted photon numbers $k$
for Alice and $l$ for Bob into arbitrary $d$-dimensional outcomes by
assigning complex variables $\omega^{k}$ and $\omega^{l}$,
respectively. Therefore, the local measurement operator for Alice is
given by
\begin{eqnarray}
\label{eq:localM}
\hat{A}(\alpha)=\sum^{\infty}_{k=0}\omega^{k}\hat{D}(\alpha)\ket{k}\bra{k}\hat{D}(-\alpha)
\equiv\omega^{\hat{k}(\alpha)},
\end{eqnarray}
and likewise for Bob $\hat{B}(\beta)\equiv\omega^{\hat{l}(\beta)}$.
The $n$th-order correlation function is then given by
\begin{eqnarray}
\nonumber \label{eq:mHOC} C^{(n)}_{\alpha\beta}&=&\langle
\hat{A}(\alpha)^{n}\hat{B}^{\dag n}(\beta)\rangle\\
&=&\sum_{k,l=0}^{\infty}\omega^{n(k-l)}P_{k,l}(\alpha,\beta),
\end{eqnarray}
where $P_{k,l}(\alpha,\beta)$ is the joint probability of counting
$k$ and $l$ photons at the local measurement setup of two modes
displaced by $\alpha$ and $\beta$, respectively. We can rewrite the
correlation function in Eq.~(\ref{eq:mHOC}) as
\begin{eqnarray}
\label{eq:hcorr}
C^{(n)}=\sum^{\infty}_{k,l=0}\biggl(\frac{s_n+1}{s_n-1}\biggr)^{k-l}
P_{k,l}(\alpha,\beta),
\end{eqnarray}
where $s_n\equiv-i\cot(n\pi/d)$. This is proportional to the
two-mode $s$-parameterized quasiprobability function
$W(\alpha,\beta;s)$ if we extend the parameter $s$ from a real to a
complex variable. The $s$-parameterized quasiprobability function is
defined as \cite{Cahill69,Moya93}
\begin{eqnarray}
\label{eq:sQP} W(\alpha; s)=\frac{2}{\pi(1-s)}\left\langle
\left(\frac{s+1}{s-1}\right)^{\hat{n}(\alpha)} \right\rangle,
\end{eqnarray}
where $\hat{n}(\alpha)$ is a number operator displaced by a complex
variable $\alpha$ in phase space. It becomes the $P$-function, the
Wigner-function, and the $Q$-function when setting $s=1, 0, -1$
\cite{Moya93}, respectively. Then the correlation function is
written by a two-mode quasiprobability function as
\begin{eqnarray}
\label{eq:corrbyqf}
C^{(n)}_{\alpha\beta}&=&\frac{\pi^2(1-s_n^2)}{4}W(\alpha, \beta;
s_n).
\end{eqnarray}
We define quasiprobability functions of, e.g., the two-mode squeezed
vacuum states given in Eq.~(\ref{eq:TMSVstate}) with complex
variable order parameter as follows. The characteristic function for
two-mode squeezed vacuum states is defined using a complex order
parameter $s_n$ by
\begin{eqnarray}
\chi(\xi,\eta;s_n)&=&_{\mathrm{TMSS}}\bra{\Psi}\hat{D}(\xi)\otimes\hat{D}(\eta)
\ket{\Psi}_{\mathrm{TMSS}}\\
\nonumber&&~~~~~~~~~~~~\times\exp(s_n|\xi|^2+s_n^*|\eta|^2/2)\\
\nonumber
&=&\exp\biggl[-\frac{1}{2}\{|\xi|^2(\cosh{2r}-s_n)\\
\nonumber
&&+|\eta|^2(\cosh{2r}-s_n^*)+(\xi\eta+\xi^*\eta^*)\sinh{2r}\}\biggr].
\end{eqnarray}
The corresponding quasiprobability functions can be obtained by
\begin{eqnarray}
\label{eq:tmssqp}
W(\alpha,\beta;s_n)&=&\frac{1}{\pi^4}\int^{\infty}_{-\infty}d^2\xi
d^2\eta~\chi(\xi,\eta;s_n)\\
\nonumber
&&~~~~~~\times\exp(\alpha\xi^*-\alpha^*\xi)\exp(\beta\eta^*-\beta^*\eta)\\
\nonumber
&=&\frac{4}{\pi^2(1-s_n^2)}\exp\biggl[-\frac{2}{1-s_n^2}\{|\alpha|^2A^{*}\\
\nonumber &&~~~~~~+|\beta|^2A+(\alpha\beta+\alpha^*\beta^*)\sinh{2r}
\}\biggr],
\end{eqnarray}
where $A=\cosh{2r}-s_n$. Therefore, from the Eqs.~(\ref{eq:hcorr}),
(\ref{eq:sQP}), and (\ref{eq:tmssqp}), we obtain the correlation
function for two-mode squeezed vacuum states as
\begin{eqnarray}
\label{eq:appencorr}
C^{(n)}_{\alpha\beta}=\exp\biggl[-2\frac{|\alpha|^2A^{*}+|\beta|^2A+
(\alpha\beta+\alpha^*\beta^*)\sinh{2r}}{1-s_n^2}\biggr].
\end{eqnarray}

Now we can rewrite two types of high-dimensional Bell inequalities
in terms of quasiprobability functions; that is the correlation
functions of CGLMP given in Eq.~(\ref{eq:CGLMP2}) and SLK in
Eq.~(\ref{eq:SLK2}) can be replaced with quasiprobability functions
using Eq.~(\ref{eq:corrbyqf}). Note that for two-dimensional
outcomes $d=2$, the correlation function is proportional to the
two-mode Wigner function,
\begin{eqnarray}
\label{eq:Wign}
C_{\alpha\beta}=\sum^{\infty}_{k=0}(-1)^{k-l}P_{k,l}(\alpha,\beta)
=\frac{\pi^2}{4}W(\alpha,\beta),
\end{eqnarray}
and in this case both, CGLMP and SLK, become equivalent with the
type proposed in \cite{Banaszek99} in the form of the CHSH Bell
inequality.

On the basis of this formalism we will investigate violations of
high-dimensional Bell inequalities, CGLMP and SLK, for any quantum
state that can be represented by the quasiprobability functions.

\subsection{Violations of Bell inequalities} \label{section:Violation}

\begin{figure}
\begin{center}
\includegraphics[width=0.49\textwidth]{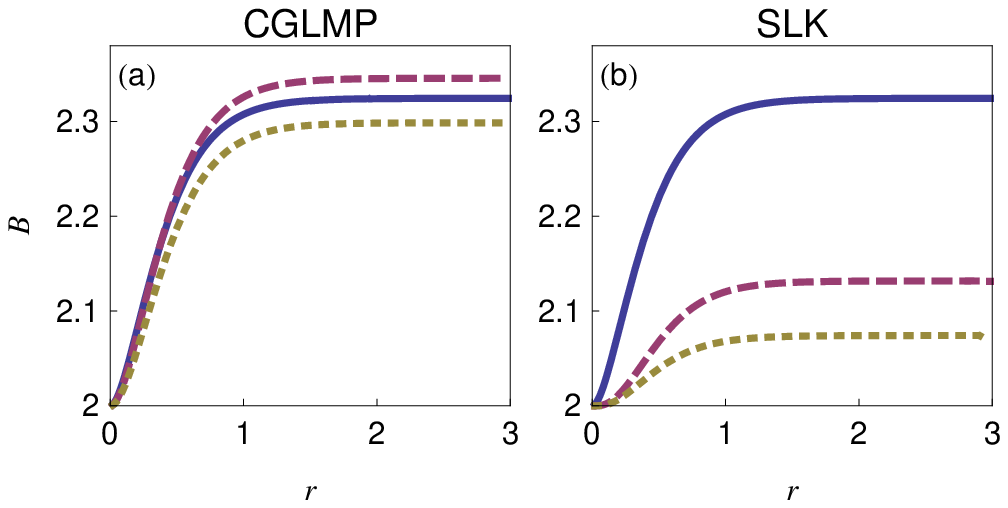}
\includegraphics[width=0.47\textwidth]{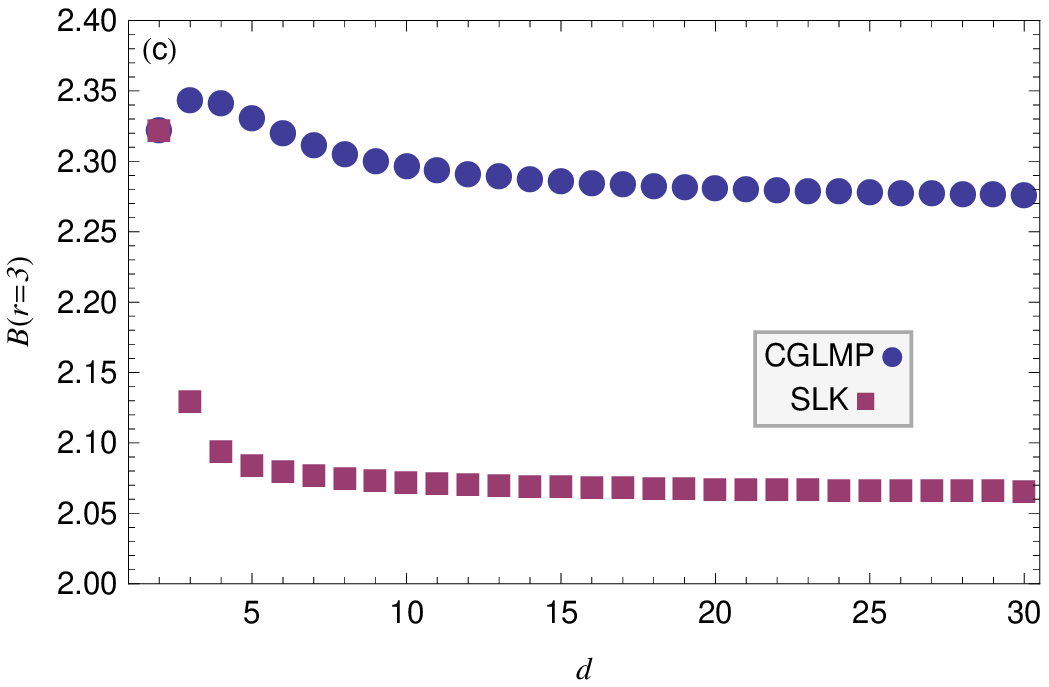}
\caption{(Color online) Violations of the (a) CGLMP and (b) SLK
inequalities for two-mode squeezed vacuum states with varying
squeezing rate $r$ when $d=2$ (solid line), $d=3$ (dashed line) and
$d=10$ (dotted line). (c) The expectation values of CGLMP and SLK
are compared while increasing the number of outcomes $d$ for a
two-mode squeezed vacuum state ($r=3$).}\label{fig1}
\end{center}
\end{figure}

We investigate violations of two types of Bell inequalities, CGLMP
and SLK, for two-mode squeezed vacuum states by properly choosing
local measurements $\alpha$, $\alpha'$, $\beta$ ,and $\beta'$. The
maximal expectation values of the CGLMP and SLK functions for
different numbers of outcomes $d$ are plotted in Figs.~\ref{fig1}(a)
and \ref{fig1}(b), respectively, against the squeezing rate $r$. The
expectation values of both types exceed the local-realistic bound
$2$ for any $r$ and increase up to maximum as $r$ increases.
However, the degrees of violation of the two types show different
tendencies depending on the number of outcomes $d$ as shown in
Fig.~\ref{fig1}(c).

For CGLMP inequalities, the degree of violation reaches a maximum
when $d=3$ and decreases as $d$ increases. Tests of the CGLMP
inequality for $d=3$, $d=4$, and $d=5$ exhibit stronger violations
than that of the CHSH inequality in agreement with the results in
Ref.~\cite{WSon06}. This is an advantage offered by the CGLMP
inequality tests over the CHSH inequality test. For $d>5$, the
degree of violation is lower than that of a two-dimensional test.
Nevertheless, the expectation value does not decrease quickly so one
can still verify strong violations of the local realism in
high-dimensional correlations. The reason that the change in maximal
expectation values with increasing $d$ does not show a monotonous
tendency is because possible operations for local measurements are
restricted by displacement operations in phase space instead of the
full SU($d$) transformation, as pointed out in \cite{WSon06}.

On the other hand, the SLK inequalities show different tendencies.
As demonstrated in Figs.~\ref{fig1}(b) and 1(c), the degree of
violation decreases as $d$ increases, in contrast to the CGLMP
inequality, so it exhibits strongest violations when $d=2$. Note
that when $d=2$, both CGLMP and SLK types are equivalent with the
CHSH inequality and their violations by two-mode squeezed vacuum
states are the same as the results obtained in \cite{Jeong03}.

\subsection{Effects of detection inefficiency} \label{section:ImDet}

In a realistic experimental setup, noise effects occur during the
measurement process, such as photon losses and dark counts. In
general, the photon number distribution measured by inefficient
detectors $\hat{P}_{m}(\eta)$ can be modeled by the generalized
Bernoulli transformation from the real number distribution
$\ket{k}\bra{k}$ \cite{Leonhardt-1997}:
\begin{eqnarray}
\label{eq:splitpovm}
\hat{P}_{m}(\eta)=\sum^{\infty}_{k=m}\binom{k}{m}(1-\eta)^{k-m}\eta^m\ket{k}\bra{k},
\end{eqnarray}
where $\eta$ is the overall detection efficiency and
$\sum^{\infty}_{m=0} \hat{P}_{m}(\eta)=\openone$. We shall not
consider dark counts here as those are relatively minor when the
detection efficiency is low. It is known that dark count rates can
be suppressed when low-efficiency detectors are used: Highly
efficient detectors have relatively high dark count rates, while
less efficient detectors have very low dark count rates
\cite{Takeuchi99}.

The realistic local measurement operator for Alice with detection
efficiency $\eta_A$ is given by
\begin{eqnarray} \label{eq:lop} \nonumber
\hat{A}(\alpha,\eta_A)&=&\sum^{\infty}_{m=0}\omega^{m}\hat{D}(\alpha)\hat{P}_{m}(\eta)\hat{D}(-\alpha)\\
\nonumber
&=&\sum^{\infty}_{k=0}(1-\eta_A+\eta_A\omega)^k\hat{D}(\alpha)\ket{k}\bra{k}\hat{D}(-\alpha)\\
&=&(1-\eta_A+\eta_A\omega)^{\hat{k}(\alpha)},
\end{eqnarray}
and likewise for Bob
$\hat{B}(\beta,\eta_B)=(1-\eta_B+\eta_B\omega)^{\hat{l}(\beta)}$.

The correlation function between Alice and Bob for two mode squeezed
states is written by
\begin{eqnarray}
\label{eq:cfnoise} \nonumber
C^{(n)}_{\alpha\beta}(\eta_A,\eta_B)=\frac{S(\eta_A,\eta_B)}{T(\eta_A,\eta_B)}\exp\biggl[-\frac{2}{T(\eta_A,\%
eta_B)}\{|\alpha|^2R^{*}(\eta_B)\\
+|\beta|^2R(\eta_A)+(\alpha\beta+\alpha^*\beta^*)\sinh{2r}\}\biggr],~~
\end{eqnarray}
where
\begin{eqnarray}
\nonumber
R(\eta)=\cosh{2r}-1+\frac{1}{\eta}+\frac{i}{\eta}\cot\frac{n\pi}{d},\\
\nonumber
S(\eta_A,\eta_B)=\frac{1}{\eta_A \eta_B}(1+\cot^2\frac{n\pi}{d}),\\
\nonumber T(\eta_A,\eta_B)=R(\eta_A)R^{*}(\eta_B)-\sinh^2{2r},
\end{eqnarray}
which becomes equivalent to Eq.~(\ref{eq:appencorr}) when
$\eta_A=\eta_B=1$. The expectation values of CGLMP and SLK in the
presence of detection inefficiency are then obtained by applying
Eq.~(\ref{eq:cfnoise}) to Eqs.~(\ref{eq:CGLMP2}) and
(\ref{eq:SLK2}), respectively. Since violations of the SLK
inequality become weaker as $d$ increases, even in the case of
perfect efficiency shown in Sec.~\ref{section:Violation}, we here
consider only the CGLMP inequality.

\begin{figure*}
\begin{center}
\includegraphics[width=0.9\linewidth]{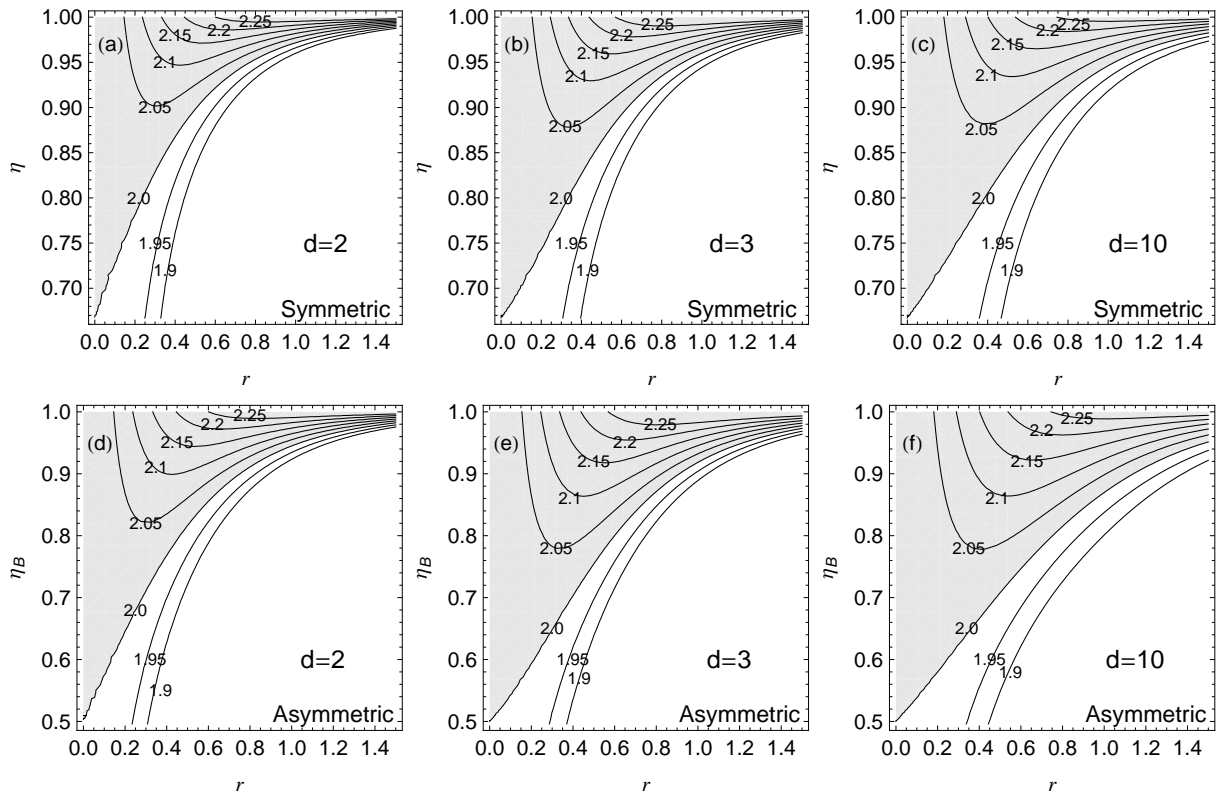}
\caption{Expectation values of the CGLMP tests for two-mode squeezed
vacuum states in symmetric cases($\eta_A=\eta_B=\eta$) for (a)
$d=2$, (b) $d=3$, and (c) $d=10$, and asymmetric cases($\eta_A=1$)
for (d) $d=2$, (e) $d=3$, and (f) $d=10$. The shaded regions
indicate the violations of the CGLMP inequalities in the range of
the detection efficiency $\eta$ and the squeezing rate
$r$.}\label{fig2}
\end{center}
\end{figure*}

Let us first consider the {\em symmetric} case when the detector
efficiencies of Alice and Bob are the same $\eta_A=\eta_B=\eta$.
Figures~\ref{fig2}(a), 2(b) and 2(c) show violations of the CGLMP
inequality when $d=2, 3$ and $10$, respectively, in the range of
efficiency $\eta$ and squeezing rate $r$. It is shown that
high-dimensional tests can exhibit stronger violations than that for
$d=2$ in some regions of $\eta$ and $r$. Furthermore, it is
noticeable that the bound efficiency for observing quantum
nonlocality becomes lower as $d$ increases for a given $r>0$. For
example, for a two-mode squeezed vacuum state $r=0.3$ and a
detection efficiency $\eta=0.8$, one can observe quantum nonlocality
when testing the CGLMP inequality with $d=10$, while one can not
observe it when testing the CHSH inequality ($d=2$). We note that
the bound efficiency for any $d$ is down to $\eta = 0.667$ as $r$
decreases to zero, which interestingly is the Eberhard limit, i.e.,
the lowest bound efficiency for the CHSH Bell test
\cite{Eberhard1993}. This is because for slightly squeezed states
the first two levels of number basis are dominant, so the CGLMP Bell
test becomes nearly equivalent to the two-dimensional test. It is
also notable that for any $d$ the efficiency bound becomes higher as
the squeezing rate $r$ increases, and thus, the violation for the
Einstein-Podolsky-Rosen (EPR) state ($r=\infty$) is observed only
when $\eta=1$. This may be because the number counting with a
displacement operation is not an optimal local measurement for
testing nonlocality with the EPR state, as pointed out in Ref.
\cite{SWLEE09-2}.

Let us also consider an {\em asymmetric} case when $\eta_A=1$ and
thus the effects of inefficiency are characterized only by $\eta_B$.
This can be realized by an atom-photon entanglement since the atom
is measured with an efficiency close to 1
\cite{Cabello2007,Brunner2007}. Figures~\ref{fig2}(d), 2(e) and 2(f)
show the violation regions of the CGLMP inequality when $d=2, 3,$
and $10$, respectively, in the range of $\eta_B$ and $r$. Similarly
to the symmetric cases, high-dimensional tests of the CGLMP
inequality are shown to be more robust to detection inefficiency
than the CHSH test for a given $r>0$. We note that the bound
efficiency for any $d$ is down to $\eta = 0.5$ as $r$ decreases to
zero, which is equivalent to the lowest limit for the CHSH Bell test
on atom-photon systems \cite{Cabello2007}.

It is shown that the CGLMP inequality offers a more robust
nonlocality test to detection inefficiency than the CHSH inequality
when using continuous-variable states. Therefore, a high-dimensional
approach may provide an advantageous way of closing the detection
loophole problem for quantum nonlocality tests, which is in
agreement with the work in Ref.~\cite{Tamas2010}.

\section{Discussion and Conclusions}
\label{section:Remarks}

The complex variable representation of correlation functions can be
efficiently used for testing high-dimensional quantum nonlocality.
Two types of high-dimensional Bell inequalities given in
Eqs.~(\ref{eq:CGLMP2}) and (\ref{eq:SLK2}) are applicable in any
case of complex-valued measurement. For example, as we have shown in
this paper, it can be extended to continuous variables by virtue of
the phase space formalism. The correlation function is then given as
a quasiprobability function with a complex order parameter, which
can be reconstructed by photon number counting.

We investigate the effect of detection inefficiency on the violation
of high-dimensional Bell inequalities when the system is given as a
pure two-mode squeezed state, while previous works have studied the
effect of system noise \cite{Collins02,Acin2002}. Similar to the
case of system noise, violations of the CGLMP inequality ($d>2$) are
shown to be more robust to detection inefficiency than that of the
CHSH inequality ($d=2$). In addition, the bound efficiency for
demonstrating quantum nonlocality becomes lower as the dimension
increases for two-mode squeezed vacuum states with a given $r>0$.
This may provide a useful insight for closing the detection loophole
problem in a nonlocality test with continuous-variable states. The
work in Ref.~\cite{SWLee09}, which shows that the $Q$-function
allows more robust Bell tests to detection inefficiency than the
Wigner function, can be understood relevantly since the $Q$-function
can be regarded as an expectation value of high-dimensional
measurements. Note that the $Q$-function is a smoothed Wigner
function where the smoothing effect is modeled as a split of
outcomes of parity measurements (i.e., $+1$ and $-1$) to
higher-dimensional outcomes.

For an experimental realization of the proposed scheme, there exists
an obstacle to overcome: the low efficiency of realistic
photon-counting detectors. As an alternative method, one may
consider a highly efficient homodyne tomography \cite{Vogel1989}.
However, for a valid quantum nonlocality test, it is required that
the quantities measured by the detectors should satisfy the
local-realistic conditions assumed when deriving the Bell
inequality. Note that the local-realistic bounds in
Eqs.~(\ref{eq:CGLMP2}) and (\ref{eq:SLK2}) are given as a maximal
expectation value of a combination of photon number correlations.
Alternatively, an atom-field interaction in a cavity can be
considered for a high-dimensional measurement \cite{Bertet02}, but
it is feasible only when the measurement dimension $d$ is a power of
2. Therefore, the realization of the proposed nonlocality test is
expected with the progress of photon detection technologies
\cite{Divochiy2008}.

In summary, we have proposed a scheme for testing high-dimensional
Bell inequalities in phase space and investigated the effect of
detection inefficiency. First, two types of high-dimensional Bell
inequalities, CGLMP and SLK, are recast into a structure composed of
complex-variable correlation functions. The correlation functions
were shown to be proportional to the quasiprobability function with
an order parameter associated with the number of outcomes, which can
be reconstructed by photon number counting. On the basis of the
proposed scheme, we demonstrated violations of two types of
high-dimensional Bell inequalities, CGLMP and SLK, for two-mode
squeezed vacuum states and compared their violations for different
numbers of outcomes. For the case of two-level outcomes, violations
of both types are equivalent to that of the CHSH inequality. For
some cases with more than two levels of outcomes the CGLMP
inequality exhibits stronger violations than the CHSH inequality,
while violation of the SLK inequality tends to decrease as the
number of outcomes increases. Finally, we have shown that the CGLMP
inequality can offer a more robust nonlocality test to detection
inefficiency than the CHSH inequality. We expect an experimental
realization of high-dimensional Bell tests on continuous-variable
states based on our scheme. An important next step will be to
increase the number of local measurement settings, which could lower
the bound efficiency further \cite{Tamas2010,Brunner2007}.

\acknowledgments

This CRI work was supported by the National Research Foundation of
Korea(NRF) funded by the Korea government(MEST) (Grant No.
3348-20100018), the Center for Subwavelenth Optics (Grant No.
R11-2008-095-01000-0), the World Class University (WCU) program, and
the TJ Park Foundation.

\end{document}